\documentclass[%
 reprint,
 amsmath,amssymb,
 aps,
]{revtex4-1}

\usepackage{graphicx}
\usepackage{caption}
\usepackage{subcaption}
\usepackage{dcolumn}
\usepackage{bm}

\usepackage{siunitx}

\begin{document}


\title{Theoretical prediction of Reynolds stresses and velocity profiles for barotropic turbulent jets}
\author{Eric Woillez}%
\email{Eric.Woillez@ens-lyon.fr}
\author{Freddy Bouchet}
\email{Freddy.Bouchet@ens-lyon.fr}
\affiliation{Univ Lyon, Ens de Lyon, Univ Claude Bernard, CNRS, Laboratoire de Physique, F-69342 Lyon, France}%

\date{\today}

\begin{abstract}
It is extremely uncommon to be able to predict the velocity profile of a turbulent flow. In two-dimensional flows, atmosphere dynamics, and plasma physics, large scale coherent jets are created through inverse energy transfers from small scales to the largest scales of the flow. We prove that in the limits of vanishing energy injection, vanishing friction, and small scale forcing, the velocity profile of a jet obeys an equation independent of the details of the forcing. We find another general relation for the maximal curvature of a jet and we give strong arguments to support the existence of an hydrodynamic instability at the point with minimal jet velocity. Those results are the first computations of Reynolds stresses and self consistent velocity profiles from the turbulent dynamics, and the first consistent analytic theory of zonal jets in barotropic turbulence.
\end{abstract}

\maketitle

Theoretical prediction of velocity profiles of inhomogeneous turbulent flows is a long standing challenge, since the nineteenth century. It involves closing hierarchy for the velocity moments, and for instance obtaining a relation between the Reynolds stress and the velocity profile. Since Boussinesq in the nineteenth century, most of the approaches so far have been either empirical or phenomenological. Even for the simple case of a three dimensional turbulent boundary layer, plausible but so far unjustified similarity arguments may be used to derive von K\'arm\'an logarithmic law for the turbulent boundary layer (see for instance ~\cite{landau2013fluid}), but the related von K\'arm\'an constant~\cite{pope2001turbulent} has never been computed theoretically. Still this problem is a crucial one and has some implications in most of scientific fields, in physics, astrophysics, climate dynamics, and engineering. Equations (\ref{eq:Reynolds stress small scale}-\ref{eq:dynamics}), (\ref{eq:diverging solution}), and (\ref{eq:curvature}) are probably the first prediction of the velocity profile for turbulent flows, and relevant for barotropic flows.

In this paper we find a way to close the hierarchy of the velocity moments, for the equation of barotropic flows with or without effect of the Coriolis force. This two dimensional model is relevant for laboratory experiments of fluid turbulence~\cite{paret1997experimental}, liquid metals~\cite{sommeria1986experimental}, plasma~\cite{dubin1988two}, and is a key toy model for understanding planetary jet formation~\cite{marston2008statistics} and basics aspects of plasma dynamics on Tokamaks in relation with drift waves and zonal flow formation~\cite{diamond2005zonal}. It is also a relevant model for Jupiter troposphere organization~\cite{galperin2014cassini}. Moreover, our approach should have future implications for more complex turbulent boundary layers, which are crucial  in climate dynamics in order to quantify momentum and energy transfers between the atmosphere and the ocean.

It has been realized since the sixties and seventies in the atmosphere dynamics and plasma communities that in some regimes two dimensional turbulent flows are strongly dominated by large scale coherent structures. Jets and large vortices are often observed in numerical simulations or in experiments, but the general mechanism leading to such an organization of the flow at large scales is subtle and far from being understood. For simplicity, we consider in this paper the case of parallel jets favored by the $\beta$ effect, however without $\beta$ effect both jets and vortices can be observed~\cite{sommeria1986experimental,simonnet,frishman2017jets}. When a large scale structure is created by the flow, a quasilinear approach may be relevant. Such a quasilinear approach requires solving a coupled equation for the mean flow and the Lyapunov equation that describes the fluctuations with a Gaussian approximation, just like the Lenard--Balescu equation in plasma kinetic theory. Numerical approaches and theoretical analysis have been systematically developed for fifteen years in order to solve and study such quasilinear or related approximations~\cite{farrell2003structural,marston2008statistics}. In a recent theoretical paper, the range of validity of such an approach has been established by proving the self consistency of the approximations of weak forcing and dissipation limit ~\cite{Bouchet_Nardini_Tangarife_2013_Kinetic_JStatPhys}. While this work gave theoretical ground to the approach, explicit formula for the Reynolds stress cannot be expected in general. However, in a recent work~\cite{laurie2014universal}, an expression for the Reynolds stress has been derived from the momentum and energy balance equations by neglecting the perturbation cubic terms in the energy balance (this follows from the quasilinear approach justification~\cite{Bouchet_Nardini_Tangarife_2013_Kinetic_JStatPhys}), but also neglecting pressure terms (not justified so far, see~\cite{falkovich2016interaction}). This approach surprisingly predicts a constant velocity profile for the outer region of a large scale vortex in two dimensions that does not depend on the detailed characteristics of the stochastic forcing but only on the total energy injection rate $\epsilon$ expressed in \si{\meter^{2}\second^{-3}}. Another analytic expression for the Reynolds stress has also been derived in the particular case of a linear velocity profile $U$ in ~\cite{srinivasan2014reynolds}. For the case of dipoles for the 2D Navier--Stokes equations, the papers ~\cite{kolokolov2016velocity,kolokolov2016structure} following a computation analogous to the one given in ~\cite{srinivasan2014reynolds}, shows that if the vorticity is passively advected (the third term in equation (5) below is neglected), then the expression for the Reynolds stress discussed in [10] is recovered (~\cite{kolokolov2016structure} also discusses other interesting aspects related to parts of the flow for which this relation is not correct). What are the criteria for the validity of these results? Can we reconcile the different results giving a full theoretical justification and extend these for more general cases?

We start from the equations for a barotropic flow on a periodic beta
plane with stochastic forcing 
{\small
\begin{equation}
\partial_{t}{\mathbf V}+{\mathbf V}.\nabla {\mathbf V}  =  -r{\mathbf V}-\frac{1}{\rho}\nabla P+\beta_{d}y\left(\begin{array}{c}
V_{y}\\
-V_{x}
\end{array}\right)+\sqrt{2\epsilon}{\mathbf f}\label{eq:navier-stokes}
\end{equation}
}where ${\mathbf V}:=\left(\begin{array}{c}
V_{x}\\
V_{y}
\end{array}\right)$ is the two dimensional velocity field with $\nabla {\mathbf V}  =  0 $. $r$ models a linear
friction, and f is a stochastic force white in time, with energy injection rate $\epsilon$, $\beta_{d}$ is the Coriolis parameter, $y$ the north-south coordinate.
Following~\cite{Bouchet_Nardini_Tangarife_2013_Kinetic_JStatPhys} we choose time and space units such that the mean kinetic energy
is 1, and $L_{x}=1$. The non dimensional equations for the vorticity $\Omega=\nabla\land{\mathbf V}$ are
{\small
\begin{equation}
\partial_{t}\Omega+{\mathbf V}.\nabla\Omega  =  -\alpha\Omega-\beta V_{y}+\sqrt{2\alpha}\eta\label{eq:Navier-stokes adim}
\end{equation}}
where $\eta=\nabla\land f$, and ${\mathbf V}$ denotes from now on the nondimensional velocity. Now $\alpha=L\sqrt{\frac{r^{3}}{\epsilon}}$
is a nondimensional parameter although we will often refer to it as
the ``friction''. $\beta=\sqrt{\frac{r}{\epsilon}}L^{2}\beta_{d}$
is the nondimensional Coriolis parameter.  Eq. (\ref{eq:Navier-stokes adim}) still has three nondimensional
parameters, $\alpha,\beta$ and $K$, the typical Fourier wavenumber
where energy is injected. 

Neglecting the pressure and cubic terms in the energy balance and enstrophy balance, it is straightforward to obtain the Reynolds
stress expression
\begin{equation}
\left\langle uv\right\rangle =\frac{\epsilon}{U'}\label{eq:Reynolds stress}
\end{equation}
where $U'=\rm{d}U/\rm{d}y$. This generalizes the result obtained for a vortex \cite{laurie2014universal} to the case of a jet with mean velocity $U$. 
Is it possible to justify those hypothesis on theoretical ground, uncover the validity range of (\ref{eq:Reynolds stress}), and to generalize it? We note that detailed numerical studies of the energy balanced has been discussed in several papers  \cite{tangarife-these,BouchetNardiniTangarife2015,frishman2017jets}.

In order to derive eq. (\ref{eq:Reynolds stress}), the key idea is to use the already justified \cite{Bouchet_Nardini_Tangarife_2013_Kinetic_JStatPhys} quasilinear approximation in the limit of small forces and friction (inertial regime, $\alpha \ll 1$), and to further consider the limit of small scale forcing ($K \gg 1$), with fixed $\beta$. In these limits, energy is injected at small scale and is dissipated at the largest scale of the flow. $\alpha \ll 1$ is the proper regime for most geophysical turbulent flows, for instance for giant gaseous planets like Jupiter \cite{porco2003cassini,salyk2006interaction}, and many two dimensional or rotating turbulence experiments. The small scale forcing limit $K\gg1$ is the most common framework for turbulence studies (see for ex. \cite{falkovich2016interaction}) and relevant for Jupiter troposphere. Also, computing the pressure from the Navier-Stokes equations involves inverting a Laplacian. It is thus natural to expect the pressure term to have a power expansion in the parameter
$\frac{1}{K}$, and thus vanish in the limit of large $K$. The main idea is then to separate the flow ${\mathbf V}$ in two parts, ${\mathbf V}({\mathbf r},t)=U(y,t){\mathbf e}_{x}+\left(\begin{array}{c}
u({\mathbf r},t)\\
v({\mathbf r},t)
\end{array}\right)$. The mean velocity $U(y){\mathbf e}_{x}=\frac{1}{L_{x}}\int{\rm d}x\mathbb{E}[{\mathbf V}(x,y)]$ called the \emph{mean flow} or \emph{zonal
flow}, is defined
as both the zonal and stochastic average of the velocity field. In the following, the bracket $\left\langle \right\rangle $
will be used for this zonal and stochastic average. We are left with two coupled equations, one governing the dynamics
of the mean flow, the other one describing the evolution of eddies.
In the limit where $\alpha$ is small, it has been proven that fluctuations
are of order $\sqrt{\alpha}$ and thus it is self-consistent to neglect
nonlinear terms in the equation for fluctuations \cite{Bouchet_Nardini_Tangarife_2013_Kinetic_JStatPhys}. Then one can justify \cite{Bouchet_Nardini_Tangarife_2013_Kinetic_JStatPhys} that, at leading order in $\alpha$, the full velocity field statistics are described by a quasi-Gaussian field (the velocity field is not Gaussian, but the marginals when the zonal flow is fixed are Gaussian, justifying a posteriori a second order closure corresponding to the quasilinear approximation). 
Using also the incompressibility condition, we obtain the quasilinear 
model
\begin{eqnarray}
&\partial_{t}U=-\alpha\left[\partial_{y}\left\langle uv\right\rangle +U\right]\label{eq:quasilinear}\\
&\partial_{t}\omega+U\partial_{x}\omega+(\beta-U'')v=-\alpha\omega+\eta\label{eq:quasilinerar2}
\end{eqnarray}
where we have introduced $\omega=\partial_{x}v-\partial_{y}u=\triangle\psi$,
the vorticity of the fluctuations. Eq. (\ref{eq:quasilinear})
shows that the typical time scale for the evolution of the mean flow
$U$ is $\frac{1}{\alpha}$ which is, following our assumption $\alpha\ll1$,
much larger than the time scale for the evolution of eddies. Using
this time scale separation, we will consider that $U$ is a constant
field in the second Eq. (\ref{eq:quasilinerar2}), and we will
always solve $\omega(t)$ for a given $U$. 
We follow the strategy: 
\begin{itemize}
\item First we solve the linear Eq. (\ref{eq:quasilinerar2}) and compute
the stationary distribution \emph{$\left\langle \omega^{2}\right\rangle $
as a functional of $U$}. 
\item The enstrophy balance for the fluctuations allows us to relate $\left\langle \omega^{2}\right\rangle $
to the divergence of the Reynolds stress tensor (see  the Supplementary Material in appendix). 
\item Last we can use this expression to close the first Eq. (\ref{eq:quasilinear}),
and discuss possible stationary profiles $U$.
\end{itemize}
To reach the first objective, we take advantage
of the asymptotic regimes $\alpha\rightarrow0$ and $K\rightarrow\infty$.
When we take those two limits, it is natural to ask whether they commute
or not, and which nondimensional parameter will govern the difference
between $\alpha\rightarrow0$ first or $K\rightarrow\infty$
first. Our asymptotic calculations show that the key parameter is the ratio between $\frac{U''}{K}$
and $\alpha$. Taking the limit $\alpha\rightarrow0$ first amounts to saying that $\frac{U''}{\alpha K}$
is very large.

We first take the limit $K\rightarrow\infty$ while keeping $\alpha$ small but finite. The idea is to write Eq.
(\ref{eq:quasilinerar2}) in an integral form using the Green function
of the Laplacian, and use the fact that the Green function decreases
very fast when $K$ is large, which implies that the evolution of
the flow is local in space. At this stage of the calculation, $\alpha$
is small but finite, and the expression for the Reynolds stress depends
both on $\alpha$ and on the properties of the stochastic forcing. The complete calculation is reported in the Supplementary Material \cite{Bouchet_Woillez_Jupiter_jets}.  
We emphasize that as long as $\alpha$ is kept finite, the Reynolds stress depends on the Fourier spectrum of the stochastic forcing $\eta$. The result shows that the Reynolds stress can be expressed
analytically as
{\small
\begin{equation}
\left\langle uv\right\rangle =\frac{1}{2\alpha}\chi\left(\frac{U'}{2\alpha}\right),\label{eq:Reynolds stress small scale}
\end{equation}}
where the explicit expression of $\chi$ is a parametric integral, see \cite{Bouchet_Woillez_Jupiter_jets}. The stationary profile $U$ thus verifies $\frac{U''}{2\alpha}\frac{1}{2\alpha}\chi'\left(\frac{U'}{2\alpha}\right)=-U$,
which can be integrated using a primitive $X$ of the function $x\rightarrow x\chi'(x)$
in
{\small 
\begin{equation}
X\left(\frac{U'}{2\alpha}\right)+\frac{1}{2}U^{2}=C,\label{eq:dynamics}
\end{equation}}
where $C$ is the integration constant.
It is in itself remarkable that for some range of the parameters,
the flow of the barotropic quasilinear model can be computed from
a Newtonian equation like (\ref{eq:dynamics}).
 
\begin{figure}
\begin{centering}
\includegraphics[scale=0.42]{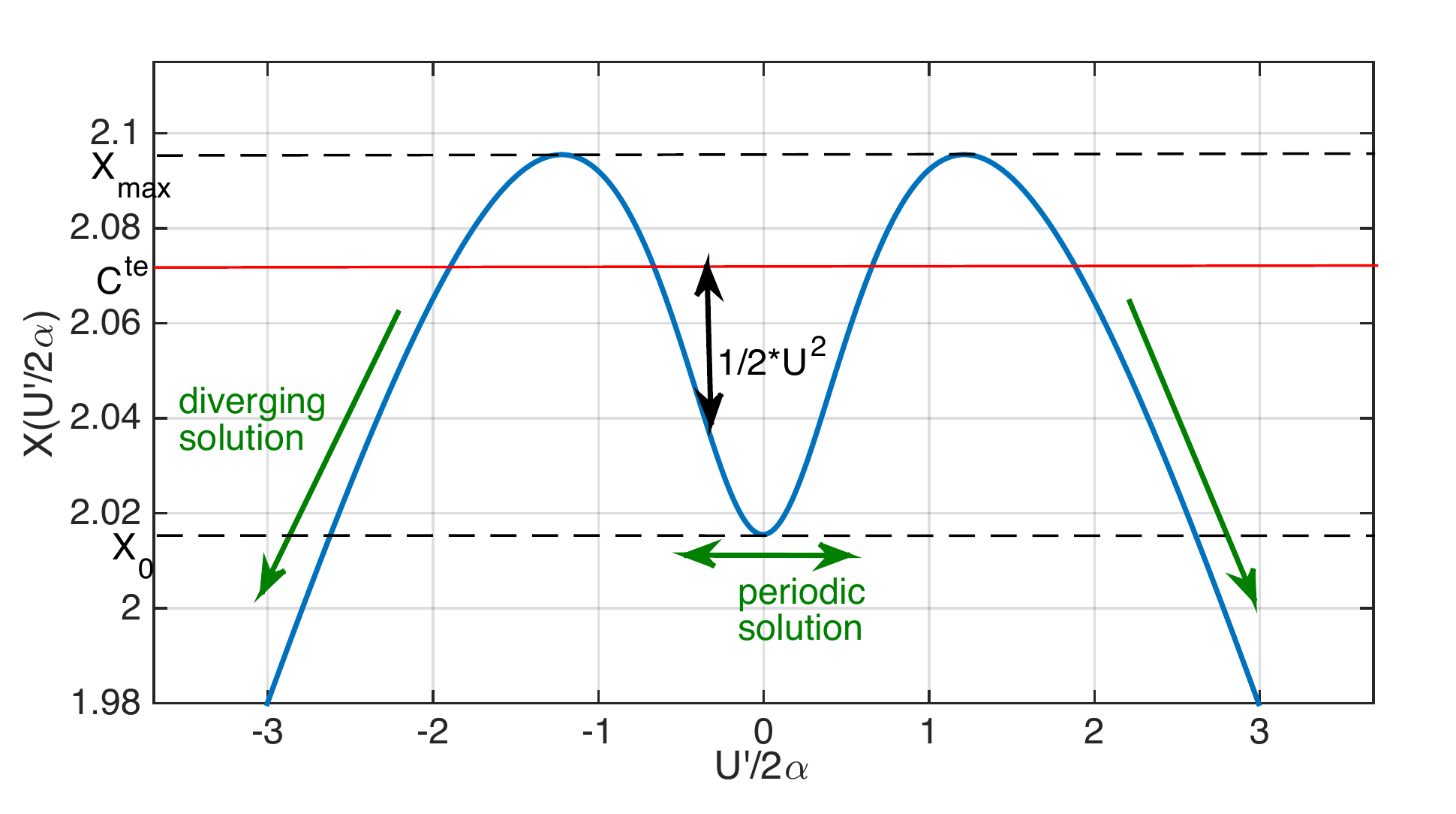}
\end{centering}
\caption{{\small In the limit of small scale forcing, the mean flow can be computed analytically from the Newtonian Eq. (\ref{eq:dynamics}). We show here that the situation is analogous to a particle moving in a one-dimensional potential. The blue curve displays the potential $X$ appearing in Eq.(\ref{eq:dynamics}). If the constant of motion is less than $X_{max}$, there are two classes of solutions: the solution can be confined in the central well and is thus regular and periodic in space, or it is outside the well, and in this latter case it diverges.} \label{fig:petitschema}}
\end{figure}

In Eq. (\ref{eq:dynamics}), $X$ plays the
role of a potential as if the equation would describe a particle moving
in a one-dimensional potential. The constant in the right-hand side is
set by $U'(0)$ and depending on the value of this constant, there can be one, two or
three solutions as shown in Fig. (\ref{fig:petitschema}).
If $C>X_{max}$, there is one solution for which $U$ never vanishes. As the total flow momentum is zero, such solutions with either $U>0$ or $U<0$ are not physical.
If $X_{0}<C<X_{max}$, there are three possible solutions, one
is periodic, the other two diverge. The periodic solution corresponds
to $\frac{U'}{2\alpha}$ confined in the well of $X$. In that case
the flow is periodic and the solution exchanges kinetic energy in
the term $\frac{1}{2}U^{2}$ with potential energy $X\left(\frac{U'}{2\alpha}\right)$.
Outside the well, the solutions are diverging, one corresponds to
an increasing $U$ and the other to a decreasing $U$. A linear stability analysis of the periodic solution of (\ref{eq:dynamics}) shows that this solution is unstable whereas the diverging solution is stable. Thus, the periodic regular solution is not a suitable candidate for the stationary mean velocity profile $U(y)$.
If we now take the limit of vanishing $\alpha$ in expression (\ref{eq:Reynolds stress small scale}),
a straightforward calculation using the explicit expression of $\chi$ (given in \cite{Bouchet_Woillez_Jupiter_jets}) allows us to recover expression (\ref{eq:Reynolds stress}) for the Reynolds stress
because $U'\left\langle uv\right\rangle \underset{\alpha\rightarrow0}{\longrightarrow}1$
which is Eq. (\ref{eq:Reynolds stress}) with dimensional units.
The physical interpretation of the limiting case
(\ref{eq:Reynolds stress}) is very enlightening. The term $U'\left\langle uv\right\rangle $ can be interpreted
as the rate of energy transferred from small scale to large scale, therefore expression (\ref{eq:Reynolds stress}) is consistent: with the
limit of large $K$, the evolution of eddies becomes local as if the perturbation only sees a region of width $\frac{1}{K}$
around itself, and thus the different parts of the flow are decoupled.
The other limit of small $\alpha$ forces the energy to go to the
largest scale to be dissipated because the dissipation at small scales
becomes negligible. 

Let us study the other limit where $\alpha$ goes to zero first. The techniques
used in this second case are very different than the previous one. We assume in this section that the linearized dynamics has no unstable modes. The calculation involves Laplace transform tools that were used in \cite{Bouchet_Morita_2010PhyD} to study the
asymptotic stability of the linearized Euler equations. In the limit
$\alpha \ll 1$, using \cite{Bouchet_Nardini_Tangarife_2013_Kinetic_JStatPhys}, we derive in the Supplementary Material \cite{Bouchet_Woillez_Jupiter_jets} the relation between the Reynolds stress divergence and the long time behavior of
a disturbance $\omega(y,0)$ carried by a mean flow $U(y)$.  This
is an old problem in hydrodynamics, one has to solve the celebrated
Rayleigh equation
{\scriptsize
\begin{equation}
\left(\frac{d^{2}}{dy^{2}}-k^{2}\right)\varphi_{\delta}(y,c)+\frac{\beta-U''(y)}{U(y)-c-i\delta}\varphi_{\delta}(y,c)=\frac{\omega(y,0)}{ik(U(y)-c-i\delta)},\label{eq:inhomogeneous rayleigh}
\end{equation}}where $\varphi_{\delta}(y,c):=\int_{0}^{\infty}{\rm d}t\psi(y,t)e^{-ik(c+i\delta)t}$
is the Laplace transform of the stream function, and $k$ is the $x$-component
of the wavevector. The Laplace transform $\varphi_{\delta}$ is well defined  for any non zero value of the real variable $\delta$ with a strictly negative product $k\delta$. $c$ has to be understood as the phase speed of the wave, and $k\delta$ is the (negative) exponential growth  rate of the wave. Involved computations are then required to give the explicit expression of the Reynolds stress. Let us just mention that the difficulty comes from the fact that we
have to take the limit $\delta\rightarrow 0$ first in Eq. (\ref{eq:inhomogeneous rayleigh})
before $K\rightarrow\infty$. Then we can relate $\omega(y,\infty)$
to the Laplace transform $\varphi_{\delta}$ taken at $\delta=0$.
The expression for the Reynolds stress in the inertial limit involves an integral with the profile $U$ in the denominator of the integrand. The integral is defined only in regions of the flow where $U'$ does not vanish, or to state it more precisely, where the parameter $\frac{KU'}{U''}$ is large. Therefore, there exists a small
region of size $\frac{1}{K}$ around the maximum where the asymptotic
expansion breaks down. In the outer region away from the extrema,
we recover the expression (\ref{eq:Reynolds stress})
in the inertial limit $\alpha\rightarrow0$ first.
For strictly monotonic velocity profiles $U$, we conclude that the
inertial limit and the small scale forcing limit commute and that
expression (\ref{eq:Reynolds stress}) is expected to be valid.

\begin{figure}
\begin{subfigure}[b]{0.3\textwidth}
        \includegraphics[scale=0.55]{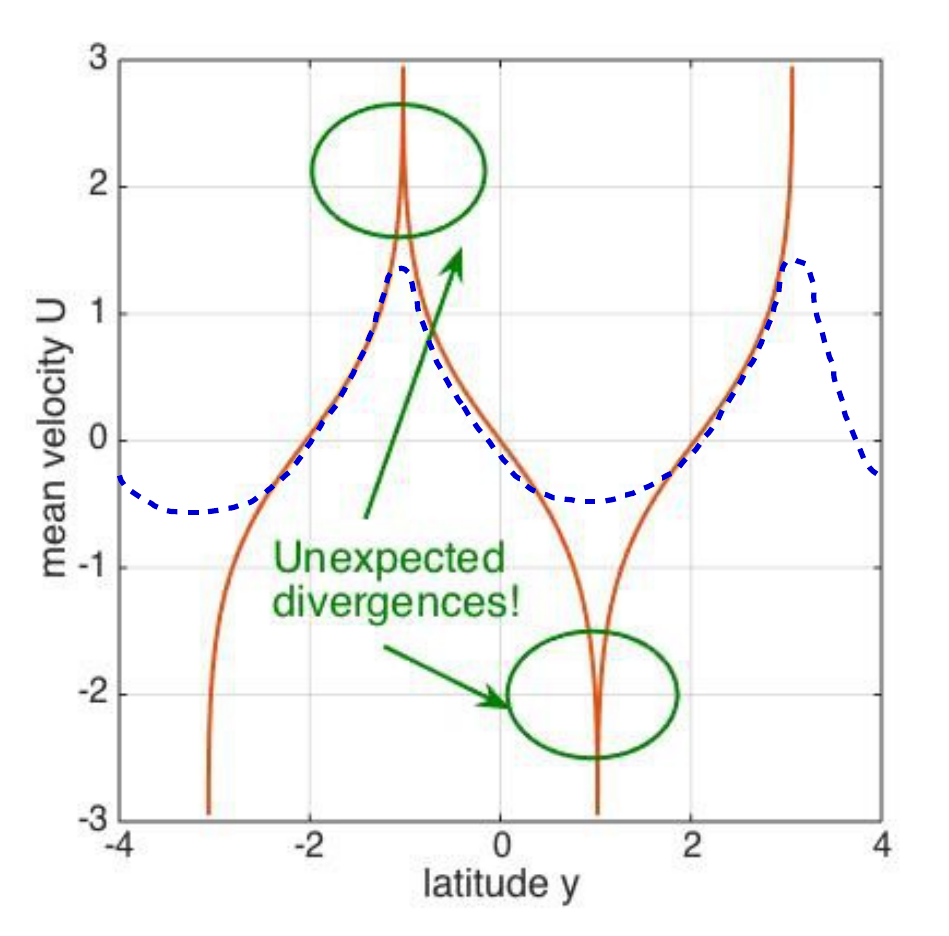}
        \caption{}
        \label{fig:divergence}
\end{subfigure}\begin{subfigure}[b]{0.19\textwidth}
        \includegraphics[scale=0.25]{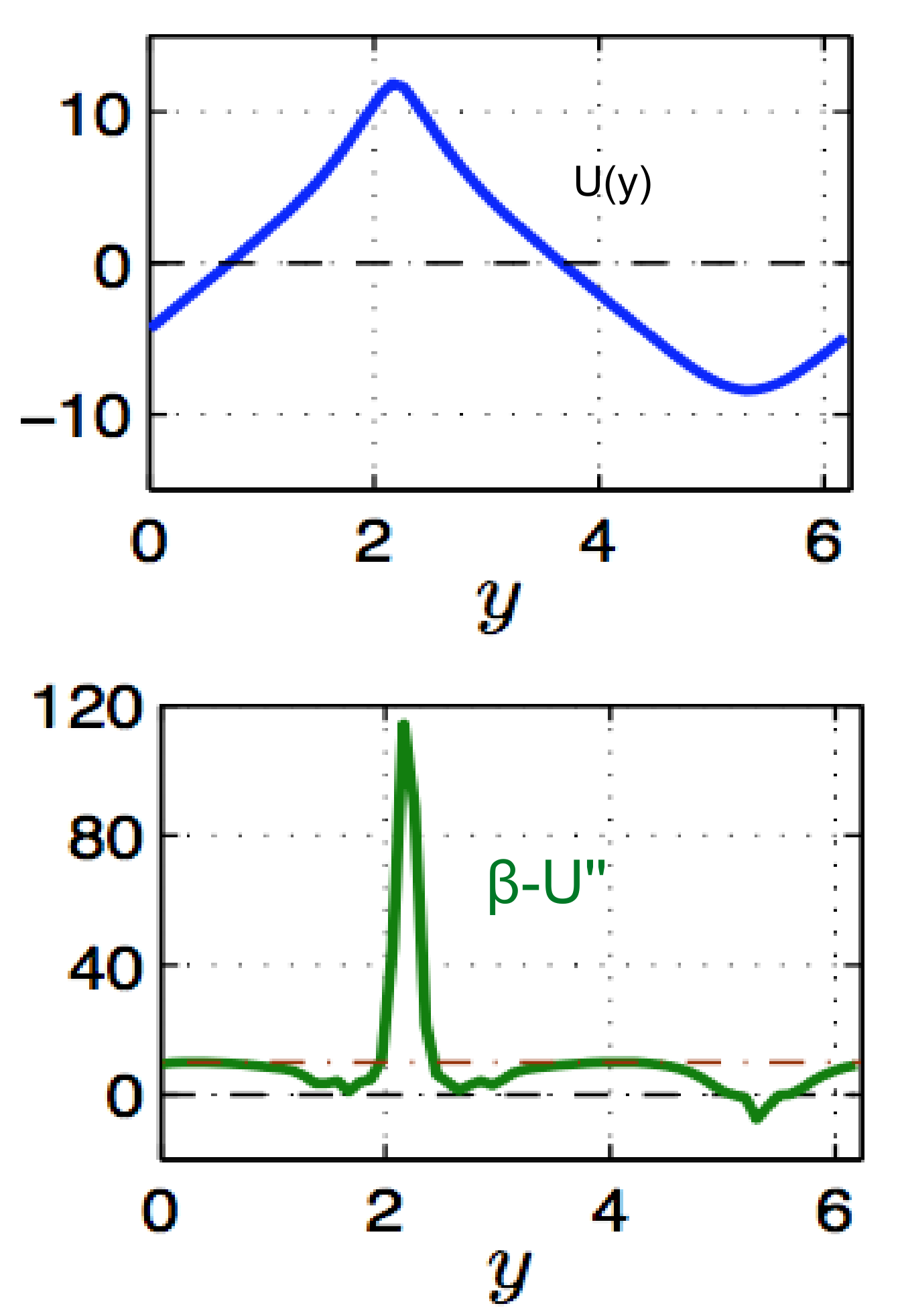}
        \caption{}
        \label{fig:numerics}
\end{subfigure}
\protect\caption{{\small (a) Jet profiles obtained as an analytic solution of (\ref{eq:diverging solution})
(red curve) together with a qualitative real velocity profile (blue
curve). (b) Real velocity profile obtained in numerical
simulations \cite{constantinou2015formation}. The analytic solution for $U$ has divergences at the extrema that are regularized by a cusp at the eastward jet and a parabola at the westward jet.}}
\end{figure}

Using expression (\ref{eq:Reynolds stress}) for the Reynolds stress,
we can solve Eq. (\ref{eq:quasilinear}) for the stationary profile.
It writes
\begin{equation}
\frac{{\rm d}}{{\rm d}y}\left(\frac{\epsilon}{U'}\right)+U=0.\label{eq:diverging solution}
\end{equation}
Whatever the value of the free parameter $U'(0)$, all profiles
$U$ are diverging in finite length. An example of such a profile
$U$ is given in Fig. (\ref{fig:divergence}). In red, we have plotted
different solutions of Eq. (\ref{eq:diverging solution}) together
such that the mean velocity profile $U$ is composed of many diverging
jets. In blue, we have drawn at hand what we expect qualitatively
from a real velocity profile. An example of a real profile obtained
by numerical simulations in \cite{constantinou2015formation} is displayed
in Fig.(\ref{fig:numerics}). The fact that Eq. (\ref{eq:diverging solution}) predicts diverging profiles $U$ shows that the expression for the Reynolds stress (\ref{eq:Reynolds stress}) is not valid everywhere in the flow, but it holds only in the spatial subdomains where the flow is monotonic, not at the extrema. We observe that both divergences are
regularized, by a cusp at the eastward jet maximum, and by a parabolic profile at the westward jet minimum. The second aim of this paper is to explain the asymmetric regularization of the eastward and westward jets.

\begin{figure}
\begin{centering}
\includegraphics[scale=0.50]{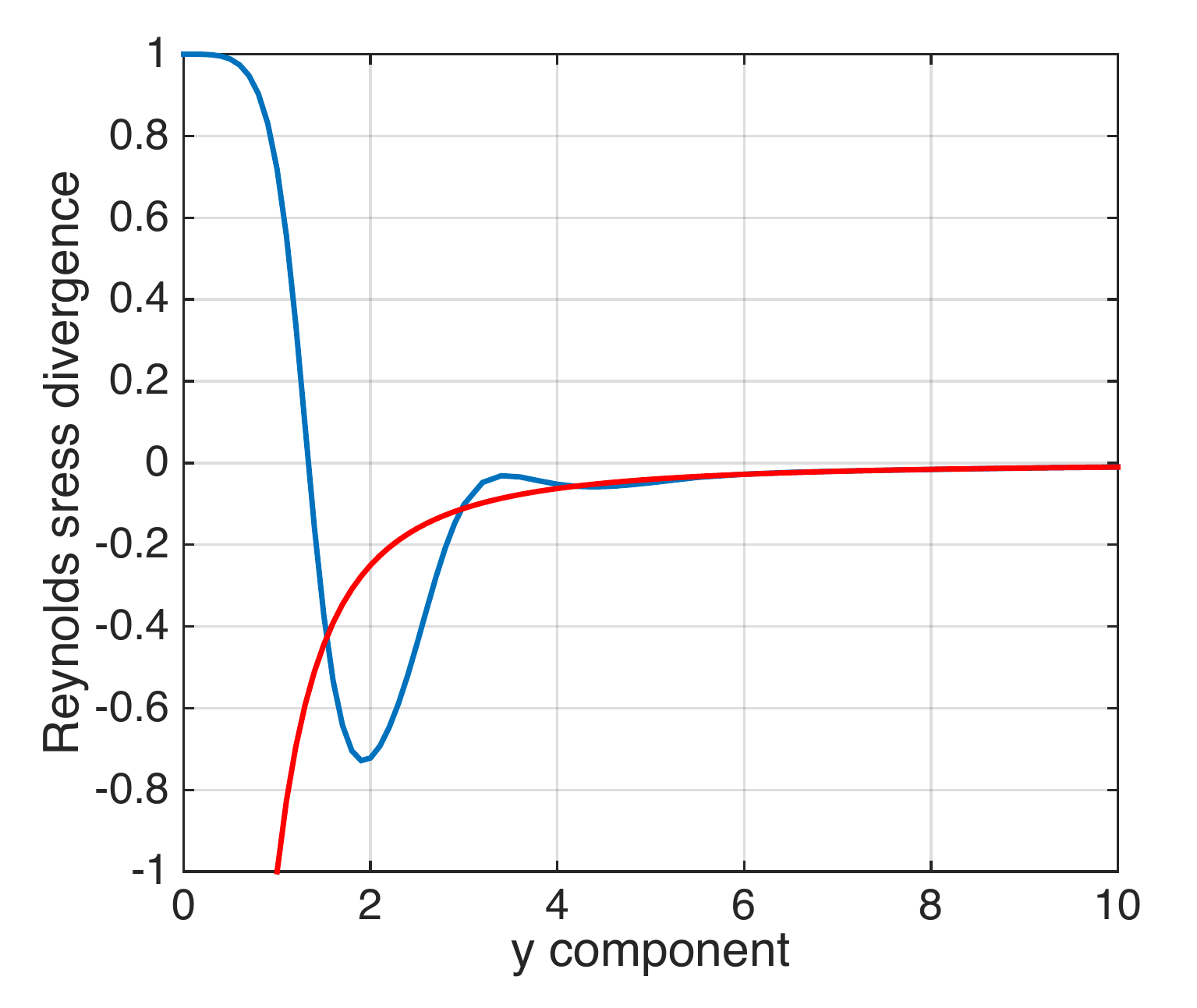}
\end{centering}

\protect\caption{{\small The Reynolds stress divergence $-\partial_{y}\left\langle uv\right\rangle $
(blue curve) computed for a parabolic profile $\widetilde{U}(y)=-\frac{y^{2}}{2}$
around its maximum $y=0$. It is satisfying to observe that the red
curve $\frac{\widetilde{U}''}{\widetilde{U}'^{2}}$ matches the blue
curve for large $y$ according to  expression (\ref{eq:Reynolds stress}).}\label{fig:tenseur cusp}}

\end{figure}

Numerical simulations like the one performed in Fig.
(\ref{fig:numerics}) show that the mean velocity profile is regularized
at a very small scale at its maximum. As we explained previously,
there exists a region of typical size $\frac{1}{K}$ around the maximum
where the asymptotic expansion for the Reynolds stress breaks down.
It is thus natural to choose the ansatz $\widetilde{U}(y):=U\left(\frac{y}{K}\right).$
The scaling in $\frac{1}{K}$ implies that the ratio $\frac{U''}{K\alpha}$
is very large at the cusp because $U''\propto K^{2}$.
The cusp is then described by the inhomogeneous
Rayleigh Eq. (\ref{eq:inhomogeneous rayleigh}). If we put the ansatz $\widetilde{U}$ in (\ref{eq:inhomogeneous rayleigh})
and consider the limit of large $K$, we get
{\scriptsize
\begin{equation}
\left(\frac{d^{2}}{dy^{2}}-\tan^{2}\theta\right)\varphi_{\delta}(y,c)-\frac{\tilde{U}"(y)}{\tilde{U}(y)-c-i\delta}\varphi_{\delta}(y,c)=\frac{e^{i\sin\theta y}}{\tilde{U}(y)-c-i\delta},\label{eq:Rayleigh cusp}
\end{equation}}where $\cos\theta=\frac{k}{K}$. The solution of this equation with
$\delta\rightarrow 0$ gives us the Laplace transform of the stream
function with which we can express the Reynolds stress. The $\beta$-effect disappears completely
from the equation of the cusp because in the region of the cusp,
the curvature is so large that it overcomes completely the $\beta$-effect,
as can be seen on the green curve of Fig. (\ref{fig:numerics}).
Eq. (\ref{eq:Rayleigh cusp}) together with the equations linking
the Reynolds stress with $\varphi$ have a numerical solution, which proves that
our scaling $\widetilde{U}(y):=U\left(\frac{y}{K}\right)$ is self-consistent.
This solution can not be expressed analytically and depends on the
Fourier spectrum of the stochastic forcing. An example of a numerical
integration of the Reynolds stress divergence $-\partial_{y}\left\langle uv\right\rangle $
for the cusp is displayed in Fig. (\ref{fig:tenseur cusp}) for
$\widetilde{U}(y)=-\frac{y^{2}}{2}$ and a stochastic forcing with
a semi annular Fourier spectrum where $\theta$ ranges between $-\frac{\pi}{3}$
and $\frac{\pi}{3}$.
In Fig. (\ref{fig:tenseur cusp}),
the red curve is the asymptote $\frac{\widetilde{U}''}{\widetilde{U}'^{2}}$
obtained from formula (\ref{eq:Reynolds stress}) with $\epsilon=1$.
When a jet is in a stationary state, the cusp profile joins smoothly
the outer region of the jet where the  result (\ref{eq:Reynolds stress})
is valid.

Another physical phenomenon at the maximum of the jet is called ``depletion
at the stationary streamlines'' and has first been observed in~\cite{Bouchet_Morita_2010PhyD}.
It means that at critical latitudes where $U'=0$, any vorticity perturbation of the flow $\omega_{0}$ has
to asymptotically vanish with time. One main consequence of this phenomenon
is the relation 
\begin{equation}
U(y_{cr})=-\frac{\epsilon K^{2}}{rU''(y_{cr})},\label{eq:curvature}
\end{equation}
where $y_{cr}$ is the latitude of the extremum. From (\ref{eq:curvature})
we learn that, even if the velocity profile of the cusp depends  on the details of the forcing,
the maximal curvature of the profile satisfies a more general relation,
and it would therefore be very interesting to check it in full numerical
simulations with different types of forcing spectra.

The previous discussion successfully explained the jet regularization of the eastward jet cusp. As clearly observed in Fig. (\ref{fig:numerics}), and from Jupiter data, westward jets do not produce cusps. At first sight, it may seem that all the theoretical arguments used so far, the energy balance, the asymptotic expansions, and the results (\ref{eq:Reynolds stress}) and (\ref{eq:curvature}), do not break the symmetry between eastward and westward jets, as $\beta$ disappears from all these computations.  However, as clearly stressed in~\cite{Bouchet_Nardini_Tangarife_2013_Kinetic_JStatPhys}, the whole theoretical approach relies on an assumption of hydrodynamic stability for $U$. The asymmetry is clearly visible in the Rayleigh-Kuo criterion, that states that when $\beta-U''$ change sign, an instability may develop. We will now argue that the turbulent flow is constantly oscillating between a stable and unstable solution, in order to control the westward jet behavior. This means that the flow is not linearly stable, but only marginally stable. As shown in the following, the instability is localized at the extremum of the westward jet, and the unstable mode has a very small spatial extension. That's why the flow can be considered as stable away from the westward extremum of the jet and the assumptions of expression (\ref{eq:Reynolds stress}) are satisfied.

\begin{figure}
\begin{centering}
\includegraphics[scale=0.43]{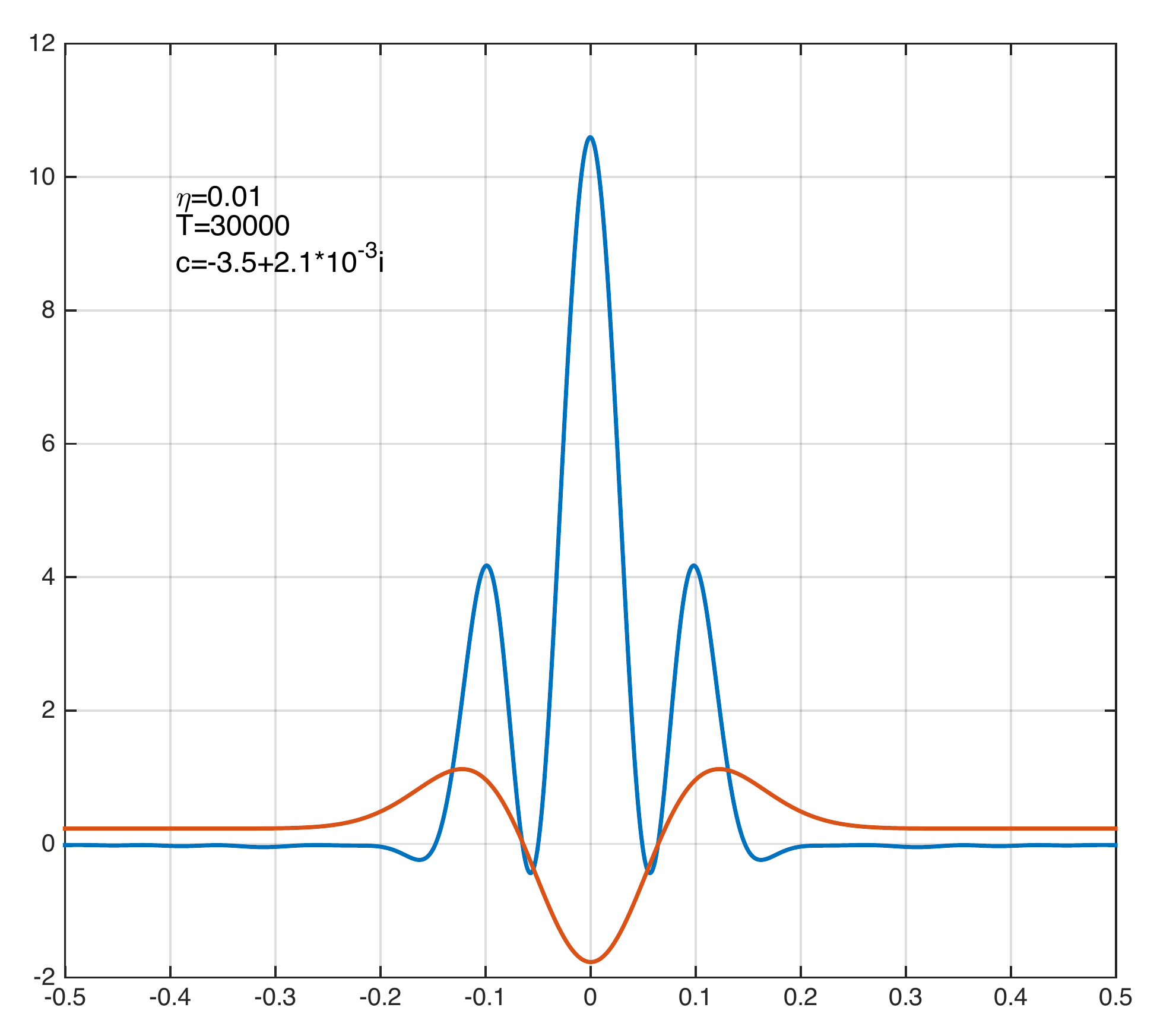}
\end{centering}

\protect\caption{Amplitude of the perturbation $|\omega^{2}(y,t)|-1$ (blue curve)
carried by a mean flow $U$. The Rayleigh-Kuo criterion $\beta-U''$
is displayed in red. The simulation shows that the perturbation grows exponentially with the complex
rate $c$. \label{fig:instabilite}}
\end{figure}

To check this marginal stability hypothesis, we solved numerically the homogeneous PDE for a perturbation
carried by a mean flow 
\begin{equation}
\partial_{t}\omega+ikU\omega+ik(\beta-U'')\psi=0.
\end{equation}
We chose a parabolic mean velocity with a small perturbation at its
minimum $U(y)=\gamma\frac{y^{2}}{2}-\eta e^{-\frac{y^{2}}{\sigma^{2}}}.$
The main curvature $\gamma$ is chosen to be slightly smaller than $\beta$,
and we performed simulations using different values of $\eta$ and
$\sigma$. 
The result of one of those simulations is displayed in
figure (\ref{fig:instabilite}). The red curve shows the Rayleigh-Kuo
criterion $\beta-U''$ which is locally violated around $y=0$. The
blue curve shows the amplitude of the perturbation $|\omega^{2}(y,t)|-1$
from which we can then express the Reynolds stress divergence.
We check that the growth rate of the perturbation is indeed exponential
with time with a complex rate $c$.

In this paper we gave a global consistent picture of how a
zonal jet is sustained in a steady state through continuous energy
transfer from small scale to large scale. Eq. (\ref{eq:dynamics}) and (\ref{eq:diverging solution}) are probably the most striking results of this work, showing for the first time that it is possible to find closed equations for the velocity profile of turbulent flows. It illustrates that, although far from equilibrium, turbulent flow velocity profiles may be described by self consistent equations, as density of other macroscopic profiles can be described in condensed matter physics. This is a fundamental property which existence is far from obvious, and that no other approach was able to establish so far. In this paper, we have considered the case of a Rayleigh friction as the mechanism for removing energy that is transferred to the largest scale. Rayleigh friction is a rather ad-hoc type of damping (although it can be justified in certain cases involving, e.g., Ekman pumping). If one would consider other kinds of friction, for instance scale-selective damping, provided that this damping actually acts on the largest scales of the flow, in a corresponding inertial limit we expect most of our results to easily generalize (the proper dissipation operator should then replace Rayleigh friction in equation (\ref{eq:quasilinear})). Indeed the processes that explain the computation of the Reynolds stress are inertial in nature and independent from the dissipation mechanisms. As an example, the case of viscous dissipation has been discussed in section 5.1.2 of 
~\cite{Bouchet_Nardini_Tangarife_2013_Kinetic_JStatPhys}, showing that while the regularization by viscosity is very different from the regularization by linear friction, the inertial results for the Reynolds stresses do coincide. This is true as far as the shear is non zero and equation (\ref{eq:Reynolds stress}) is concerned; by contrast the regularization of the cusp is dissipation dependent. In some specific cases, changing the dissipation mechanism may induce specific instability modes, like boundary layer modes due to viscous dissipation, however such effects are expected to be non generic. Extension and generalizations of our approach can be foreseen for other geometries (on the sphere), for more comprehensive quasi-geostrophic models of atmosphere jets, and for classes of flows dominated by a strong mean jet, for instance in some instances of boundary layer theory.

\acknowledgments{
We thank P. Ioannou for interesting discussion during the preliminary
stage of this work. The research leading to these results has received
funding from the European Research Council under the European Union's
seventh Framework Program (FP7/2007-2013 Grant Agreement No. 616811).}

\appendix

\section{Computation of the Reynolds stress in the small scale forcing limit\label{sec:appendixA}}

This section is devoted to the computation of the Reynold's stress
$\left\langle uv\right\rangle $ using the stochastic equation for
the fluctuations
\begin{equation}
\partial_{t}\omega+U\partial_{x}\omega+(\beta-U")v=-\alpha\omega+\eta,\label{eq:fluctuations}
\end{equation}
which is equation (5) in the article. The noise $\eta$ is white in time and has spatial correlation function $C$. Using the Ito convention for
stochastic calculus, it is straightforward to obtain the enstrophy
balance for the fluctuations
\[
\frac{1}{2}\partial_{t}\left\langle \omega^{2}\right\rangle +(\beta-U'')\left\langle v\omega\right\rangle =-\alpha\left\langle \omega^{2}\right\rangle +\frac{1}{2}C(0),
\]
and thus in statistical stationary state we get the very useful relation
\[
\left\langle v\omega\right\rangle =\frac{1}{U''-\beta}\left[\alpha\left\langle \omega^{2}\right\rangle -\frac{1}{2}C(0)\right].
\]
Thanks to the incompressibility condition, $\left\langle v\omega\right\rangle $
is related to the Reynolds stress through $\left\langle v\omega\right\rangle =-\partial_{y}\left\langle uv\right\rangle $.

First we can take advantage of the invariance along the $x$ direction
by taking the Fourier transform of (\ref{eq:fluctuations}) in $x$.
The Fourier transform in $y$ does not provide an obvious simplification
as the profile $U$ depends on $y$. However, we can use the linearity
to express the solution as the sum of particular solutions for independent
stochastic forcings $\eta_{l}(y,t)$. Each of these forcings has a
correlation function $c_{l}(y)=e^{ily}$, this means that we have
the relation\textbf{\emph{ $\mathbb{E}\left[\eta_{l}(y,t)\eta_{l}(y',t)\right]=e^{il(y-y')}\delta(t-t')$}}.
We take the Fourier transform in $x$ defined by $\omega_{k}(y):=\frac{1}{L_{x}}\int{\rm d}x\omega(x,y)e^{-ikx}$
with $k$ taking the values $\frac{2\pi}{L_{x}}n$, $n$ is an integer.
$\omega_{k,l}(y,t)$ is then defined as the function $\omega_{k}(y,t)$
that is solution of (\ref{eq:fluctuations}) with a stochastic forcing
with only one Fourier component $(k,l)$. We then obtain 
\begin{equation}
\left\langle v\omega\right\rangle =\frac{1}{U''-\beta}\underset{k,l}{\sum}\frac{\hat{C}_{k,l}}{2}\left[2\alpha\left\langle |\omega_{k,l}|^{2}\right\rangle -1\right],\label{eq:tenseur decompose}
\end{equation}
where the positive constants $\hat{C}_{k,l}$ are the Fourier coefficients
of the spatial correlation function of the noise $\eta$, $\mathbb{E}\left[\eta(x,y,t)\eta(x',y',t')\right]=C(x-x',y-y')\delta(t-t').$
Be careful that\textbf{ }in this formula the bracket $\left\langle |\omega_{k,l}|^{2}\right\rangle $
denotes a stochastic averaging, because the zonal average is already
taken into account by the sum over all vector $k$. The vorticity
$\omega_{k,l}(y,t)$ is the solution of the stochastic partial differential
equation
\begin{equation}
\partial_{t}\omega_{k,l}+ikU\omega_{k,l}+ik(\beta-U")\psi_{k,l}=-\alpha\omega_{k,l}+\eta_{l}.\label{eq:quasilinearcomplex}
\end{equation}
$\psi$ is the stream function defined through $\triangle\psi=\omega$.
As the reader would have notice, we try to reduce the problem by expressing
the solution as the sum of particular problems that we hope to be
much simpler. Now we have to find an expression for $\omega_{k,l}$
instead of the full solution $\omega$. We will go one step further
and show that the stochastic problem described by the two equations
(\ref{eq:tenseur decompose}-\ref{eq:quasilinearcomplex}) reduces
in fact to a \emph{deterministic one, }following\emph{ }\cite{Bouchet_Nardini_Tangarife_2013_Kinetic_JStatPhys}\emph{.
}Equation (\ref{eq:quasilinearcomplex}) can be formally written as
\[
\partial_{t}\omega_{k,l}+L_{k}[\omega_{k,l}]=-\alpha\omega_{k,l}+\eta_{l},
\]
where 
\begin{equation}
L_{k}[\omega_{k,l}]=ikU\omega_{k,l}+ik(\beta-U")\psi_{k,l}\label{eq:Lk}
\end{equation}
is a linear operator for a given $U$. Then we use the fact that the
noise $\eta_{k,l}$ is white in time and has an exponential correlation
function \textbf{$c_{l}(y)=e^{ily}$} to express the quantity $\left\langle |\omega_{k,l}|^{2}\right\rangle $
as
\begin{equation}
\left\langle |\omega_{k,l}|^{2}\right\rangle =\int_{-\infty}^{0}{\rm d}t~e^{2\alpha t}\left|e^{tL_{k}}[c_{l}]\right|^{2}.\label{eq:omegacarre}
\end{equation}
This formula should be understood as follows: $e^{-tL_{k}}[c_{l}]$
is the solution at time $t$ of the \emph{deterministic equation }$\partial_{t}\omega_{d}+L_{k}[\omega_{d}]=0$
with initial condition $c_{l}:=y\rightarrow e^{ily}$. The subscript
$d$ will mean that we are dealing with the solution of a deterministic
equation. The exponential $e^{2\alpha t}$ ensures the convergence
of this integral. The great advantage to have reduced the stochastic
problem to a deterministic one is that we now have to solve an hydrodynamic
problem, the propagation of a vorticity fluctuation in a shear flow,
a problem for which much has already been done in the literature.

We thus have to solve the deterministic equation 
\begin{equation}
\partial_{t}\omega_{d}+ikU\omega_{d}+ik(\beta-U'')\psi_{d}=0,\label{eq:deterministic}
\end{equation}
and we will take advantage of the small scale forcing limit where
$K\rightarrow\infty$. We will consider the case of an infinite space
in the $y$ direction, but we expect the result to be qualitatively
similar for a finite space because it is sufficient that the scale
of the forcing $\frac{1}{K}$ is small compared to the domain size,
$\frac{1}{K}\ll L_{y}.$ Please be careful here that $K$ and $k$
are different, $K^{2}=k^{2}+l^{2}$. We consider the limit $K$ infinite
with both $k$ and $l$ going to infinity\textbf{ }such that $l=k\tan\theta$
with $\theta$ a fixed constant. We will check that the limit $K\rightarrow\infty$
and $k\rightarrow\infty$ are not at all equivalent. As the leading
order term in equation (\ref{eq:deterministic}) is the free transport
term \textbf{$ikU\omega$,} we will use the natural ansatz $\omega_{d}(y,t)=A_{k,l}(y,t)e^{-ikU(y)t}$
in equation (\ref{eq:deterministic}). We obtain for $A_{k,l}$ the
equation
{\scriptsize 
\begin{equation}
\partial_{t}A_{k,l}=-i\frac{U"-\beta}{k}\int{\rm d}Y\,H_{0}(Y)A_{k,l}\left(y-\frac{Y}{k},t\right)e^{ik\left(U(y)-U(y-\frac{Y}{k})\right)t},\label{eq: amplitude}
\end{equation}}
where $H_{0}$ is the Green function of the Laplacian operator
$\left(\frac{\partial^{2}}{\partial y^{2}}-1\right)$ solving $\left(\frac{\partial^{2}}{\partial y^{2}}-1\right)H_{0}(y)=\delta(y)$.
The Green function for the full Laplacian $\left(\frac{\partial^{2}}{\partial y^{2}}-k^{2}\right)$
writes $H_{k}(y)=\frac{1}{2|k|}H_{0}(|ky|)$. For infinite space,
we have thus $H_{0}(y)=\frac{1}{2}e^{-|y|}$. This expression is very
convenient to do asymptotic calculations because, as the reader can
see, $\partial_{t}A_{k,l}$ is of order $\frac{1}{k}$ and thus goes
to zero for large $k$. At zero order, we have only free transport
of the perturbation, and the expression of $\omega$ is given by $\omega(y,t)=\omega(y,0)e^{-ikU(y)t}.$
The computation of the next order is more technical. From equation
(\ref{eq:omegacarre}) we have the basic equality

\[
2\alpha\left\langle \left|\omega_{k,l}\right|{}^{2}\right\rangle =2\alpha\int_{0}^{+\infty}e^{-2\alpha t}|A_{k,l}(y,t)|^{2}{\rm d}t.
\]
Again, the bracket $\left\langle \right\rangle $ is simply a stochastic
averaging. An integration by parts gives 
{\scriptsize
\[
2\alpha\left\langle |\omega_{k,l}|^{2}\right\rangle =|A_{k,l}(y,0)|^{2}+2\mathcal{R}e\int_{0}^{+\infty}e^{-2\alpha t}A_{k,l}(y,t)^{*}\partial_{t}A_{k,l}(y,t){\rm d}t.
\]}
Because $|A_{k,l}(y,0)|^{2}=1$, it compensates exactly the $-1$
coming from the enstrophy injection in expression (\ref{eq:tenseur decompose}).
Replacing $\partial_{t}A_{k,l}$ using (\ref{eq: amplitude}) and
equation (\ref{eq:tenseur decompose}) that gives the expression of
$\mathcal{R}e\left\langle v_{k,l}\omega_{k,l}*\right\rangle $, we
get the \emph{exact} expression for the Reynolds stress divergence
\begin{eqnarray*}
&\mathcal{R}e\left\langle v_{k,l}\omega_{k,l}^{*}\right\rangle =\frac{\hat{C}_{k,l}}{k}\mathcal{I}m\int_{0}^{+\infty}{\rm d}t...\\
&...\int{\rm d}Y\,H_{0}(Y)A_{k,l}^{*}(y,t)A_{k,l}\left(y-\frac{Y}{k},t\right)e^{ik\left(U(y)-U(y-\frac{Y}{k})\right)t}e^{-2\alpha t}.
\end{eqnarray*}
Now we will have to use our hypothesis $K\rightarrow\infty$ to go
on in the computation. Here comes a small subtlety, the $l$ component
does not appears explicitly, it is hidden in the initial condition
$A_{k,l}(y,0)=e^{ily}.$ Therefore we cannot just expand $A_{k,l}\left(y-\frac{Y}{k},t\right)$
in power of $\frac{1}{k}$ as one\textbf{ }may guess at first sight.
But expression (\ref{eq: amplitude}) tells us that $\partial_{t}A_{k,l}\rightarrow0$
as $K$ goes to infinity. The right way to do the asymptotic expansion
consists in an expansion of $A_{k,l}(y,t)$ wrt time and we use that
each temporal derivation of $A_{k,l}$ is smaller of order $\frac{1}{K}$
 
\begin{eqnarray*}
&A_{k,l}^{*}(y,t)A_{k,l}\left(y-\frac{Y}{k},t\right)=A_{k,l}^{*}(y,0)A_{k,l}\left(y-\frac{Y}{k},0\right)...\\
&...+t\partial_{t}\left[A_{k,l}^{*}(y,t)A_{k,l}\left(y-\frac{Y}{k},t\right)\right]_{t=0}+O\left(\frac{1}{K^{2}}\right),
\end{eqnarray*}
and we expand also $U$ using that $k$ is large
\begin{eqnarray*}
k\left(U(y)-U(y-\frac{Y}{k})\right) & = & U'(y)Y-\frac{U''(y)}{2k}Y^{2}\\
 & := & aY-bY^{2}.
\end{eqnarray*}
The last expression defines $a$ and $b$. Using that $A_{k,l}(y,0)=e^{ily}$
, $H_{0}(Y)=\frac{1}{2}e^{-|Y|}$ for infinite space, the first nonzero
contribution is given by
{\scriptsize
\begin{equation}
\mathcal{R}e\left\langle v_{k,l}\omega_{k,l}^{*}\right\rangle \underset{K\uparrow\infty}{\sim}\frac{\hat{C}_{k,l}}{2k}\mathcal{I}m\iint{\rm d}Y{\rm d}t\,e^{-|Y|}e^{-iY\tan\theta}e^{i\left(aY-bY^{2}\right)t-2\alpha t},\label{eq:machinappendiceB}
\end{equation}}
with $\tan\theta=\frac{l}{k}$. One could check easily that the expression
(\ref{eq:machinappendiceB}) is zero when $b=0$, that's why we have
to keep the $b$ term to get the leading order term in the expansion
in powers of $1/K$. 

We now prove\textbf{ }that the second term in the expansion of $A_{k,l}^{*}(y,t)A_{k,l}\left(y-\frac{Y}{k},t\right)$
is zero. With expression (\ref{eq: amplitude}), we compute
\[
\partial_{t}A_{k,l}(y,0)=-i\frac{U''-\beta}{k}\frac{e^{ily}}{1+\tan^{2}\theta},
\]
and then
\begin{eqnarray*}
A_{k,l}^{*}(y,0)\partial_{t}A_{k,l}\left(y-\frac{Y}{k},0\right) & = & -i\frac{U''-\beta}{k}\frac{e^{-il\frac{Y}{k}}}{1+\tan^{2}\theta}\\
\partial_{t}A_{k,l}^{*}(y,0)A_{k,l}\left(y-\frac{Y}{k},0\right) & = & i\frac{U''-\beta}{k}\frac{e^{-il\frac{Y}{k}}}{1+\tan^{2}\theta}
\end{eqnarray*}
The sum of both terms is zero. The next contribution then implies
two derivatives and is of order $\frac{1}{K^{2}}$.

The presence of both exponentials in (\ref{eq:machinappendiceB})
allows us to invert the order of integration and integrate the time
first. We get
\begin{eqnarray*}
\mathcal{R}e\left\langle v_{k,l}\omega_{k,l}^{*}\right\rangle  & \underset{K\uparrow\infty}{\sim} & \frac{\hat{C}_{k,l}}{2k}\mathcal{I}m\int{\rm d}Y\,\frac{e^{-|Y|}e^{-iY\tan\theta}}{2\alpha-i(aY-bY^{2})}.
\end{eqnarray*}
Now we have to use that $b:=\frac{U''}{2k}$ is small and develop
the denominator. This gives 
\begin{eqnarray*}
\mathcal{R}e\left\langle v_{k,l}\omega_{k,l}^{*}\right\rangle  & \underset{K\uparrow\infty}{\sim} & \frac{\hat{C}_{k,l}}{2k}\mathcal{I}m\int{\rm d}Y\frac{e^{-|Y|}e^{-iY\tan\theta}}{2\alpha-iaY+ibY^{2}}\\
 & \underset{K\uparrow\infty}{\sim} & -\frac{\hat{C}_{k,l}}{2k}\mathcal{I}m\int{\rm d}Y\,bY^{2}\frac{\partial}{\partial(aY)}\left\{ \frac{e^{-|Y|}e^{-iY\tan\theta}}{2\alpha-iaY}\right\} .
\end{eqnarray*}
 Now comes a trick: $U''Y\frac{\partial}{\partial(U'Y)}=Y\frac{\partial(U'(y)Y)}{\partial y}\frac{\partial}{\partial(U'(y)Y)}=Y\frac{\partial}{\partial y}$
and, recalling that\textbf{ $b=U''/2k$,} we can put the derivative
in $y$ in front of the integral. This is a method to recognize that
our expression is a derivative in $y$ and get the analytic expression
of the Reynolds stress. The idea to recognize a derivative seems to
fall from nowhere, but it comes in fact from the work \cite{srinivasan2014reynolds}
and the expression they derived for the Reynolds stress. In Fourier
space, the incompressibility condition writes \textbf{$\mathcal{R}e\left\langle v_{k,l}\omega_{k,l}^{*}\right\rangle =-\partial_{y}\mathcal{R}e\left\langle u_{k,l}v_{k,l}^{*}\right\rangle $,
}we then get
\[
-\partial_{y}\mathcal{R}e\left\langle u_{k,l}v_{k,l}^{*}\right\rangle \underset{K\uparrow\infty}{\sim}-\frac{\hat{C}_{k,l}}{4k^{2}}\frac{\partial}{\partial y}\mathcal{I}m\int{\rm d}Y\frac{Ye^{-|Y|}e^{-iY\tan\theta}}{2\alpha-iaY}
\]
and finally 
\[
\mathcal{R}e\left\langle u_{k,l}v_{k,l}^{*}\right\rangle \underset{K\uparrow\infty}{\sim}C^{te}+\frac{\hat{C}_{k,l}}{4k^{2}}\mathcal{I}m\int{\rm d}Y\frac{Ye^{-|Y|}e^{-iY\tan\theta}}{2\alpha-iaY}.
\]
The integration procedure defines $\left\langle uv\right\rangle $
up to a constant. This constant has no influence on the solution of
the equation for the mean velocity profile, equation (4) in the article.
In general, the expression of the Reynolds stress will depend on the
boundary conditions.

\begin{figure}
\begin{centering}
\includegraphics[scale=0.35]{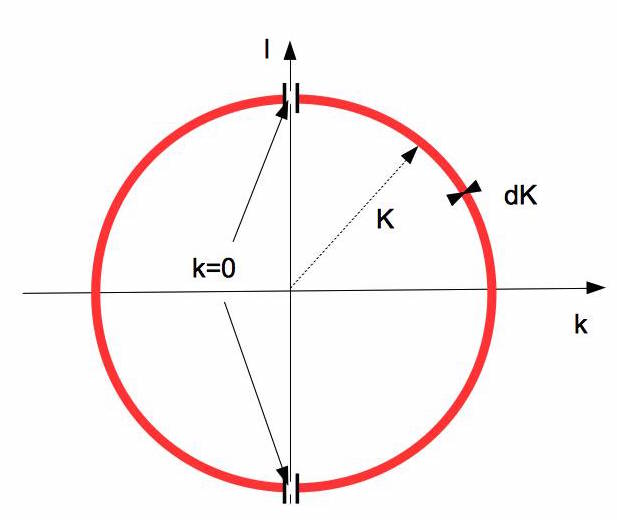}
\par\end{centering}

\protect\caption{Typical spectrum of the stochastic forcing{\label{fig:Typical-spectrum}}}
\end{figure}

Let us comment on this result. The parameter $\alpha$ is small because
we have done the quasilinear approximation, but we keep it finite.
It has a regularizing effect, and ensures that the integral remains
well defined for every profile $U$ even if $U'=0$ somewhere in the
flow. The result above does not really makes sense for $\theta$ close
to $\frac{\pi}{2}$ but the value $\frac{\pi}{2}$ is excluded because
we assumed that the stochastic forcing does not inject energy in the
mean flow. In numerical simulations, the spectrum of the stochastic
forcing is often an annulus with a weight function to make it more
or less anisotropic. The Fourier component $k=0$ is excluded because
it corresponds to a direct stochastic forcing on the zonal flow (see
figure (\ref{fig:Typical-spectrum})). To get results we can compare
to simulations, we have to integrate the contributions to the Reynolds
stress over the whole spectrum.

We will thus define a weight function $\hat{g}(\tan\theta)$ such
that the correlation function Fourier component is $\hat{C}_{k',l'}=\hat{g}(\tan\theta)K\delta(K'-K)$\textbf{\emph{.
}}On figure (\ref{fig:Typical-spectrum}) for example, we have a constant
function $\hat{g}$ except around the excluded values $\frac{\pi}{2}$
and $-\frac{\pi}{2}$. We approximate the discrete sum in (\ref{eq:tenseur decompose})
by an integral assuming $\frac{1}{L_{x}}\ll{\rm d}K.$ Because we
have fixed the mean kinetic energy to $1$, the function $\hat{g}$
must satisfy 
\begin{eqnarray*}
\frac{1}{2}\iint{\rm d}k'{\rm d}l'\frac{\hat{C}_{k',l'}}{K'^{2}}&=&\frac{1}{2}\iint K'{\rm d}K'{\rm d}\theta\frac{\hat{g}(\tan\theta)K\delta(K'-K)}{K'^{2}}\\
&=&\int_{-\infty}^{+\infty}{\rm d}x\frac{\hat{g}(x)}{1+x^{2}}\\
&=&1,
\end{eqnarray*}
In the last integral, we have done the change of variable $tan\theta=x.$
Using this function to characterize the spectrum, we obtain the simple
expression for the Reynold's stress
\begin{equation}
\left\langle uv\right\rangle (y)=\pi{\rm Im}\int_{-\infty}^{+\infty}\frac{Ye^{-|Y|}g(-Y)}{(2\alpha-iU'(y)Y)}{\rm d}Y,\label{eq:superformule2}
\end{equation}
where $g$ is the inverse Fourier transform of $\hat{g}$, $g(Y)=\frac{1}{2\pi}\int{\rm d}x\hat{g}(x)e^{iYx}$.
Now that we have an expression for the Reynolds stress in terms of
the mean profile $U$ we can solve the equation for the mean velocity
profile and find the stationary solution, if any. We let $\chi$ be
the function defined through $\chi(x):=\pi{\rm Im}\int_{-\infty}^{+\infty}\frac{Ye^{-|Y|}g(-Y)}{(1-ixY)}{\rm d}Y$,
and the stationary profile is defined by the set
\begin{eqnarray*}
\partial_{y}\left\langle uv\right\rangle  & = & -U\\
\left\langle uv\right\rangle  & = & \frac{1}{2\alpha}\chi\left(\frac{U'}{2\alpha}\right).
\end{eqnarray*}
The beautiful fact is that this system is integrable. If we replace
$\left\langle uv\right\rangle $ in the first equation, it comes
\begin{equation}
\frac{U''}{2\alpha}\frac{1}{2\alpha}\chi'\left(\frac{U'}{2\alpha}\right)=-U,\label{eq:dynamics_meanvelocity}
\end{equation}
 so we can now multiply the equality by $U'$ and integrate. Let $X$
be a primitive of $x\chi'(x)$ and we have 
\begin{equation}
X\left(\frac{U'}{2\alpha}\right)+\frac{1}{2}U^{2}=C^{te},
\end{equation}
this is equation \ref{eq:dynamics} in the article.

\section{Expression of the Reynolds stress in the inertial limit\label{sec:appendixB}}

The aim of this section is to explain how we can compute the Reynolds
stress in the inertial limit $\alpha\rightarrow0$ starting from the
relation (\ref{eq:tenseur decompose}). We have to compute
\begin{equation}
2\alpha\left\langle |\omega_{k,l}|^{2}\right\rangle =2\alpha\int_{-\infty}^{0}{\rm d}t~e^{2\alpha t}\left|e^{tL_{k}}[c_{l}]\right|^{2},\label{eq:formuleannexeA}
\end{equation}
where $e^{-tL_{k}}[c_{l}]:=\omega_{d}$ is the solution to the deterministic
equation
\begin{eqnarray}
\partial_{t}\omega_{d}+ikU\omega_{d}+ik(\beta-U")\psi_{d} & = & 0\label{eq:perturbationannexeA}\\
\left(\partial_{y}^{2}-k^{2}\right)\psi_{d} & = & \omega_{d}\nonumber 
\end{eqnarray}
with initial condition $c_{l}(y)=e^{ily}$. We will first assume there
are no neutral modes solutions of (\ref{eq:perturbationannexeA}).
First, we do the change of time scale $2\alpha t\rightarrow t$ in
the integral of (\ref{eq:formuleannexeA}). It gives us
\[
2\alpha\left\langle |\omega|^{2}\right\rangle =\int_{-\infty}^{0}{\rm d}t~e^{t}\left|e^{\frac{t}{2\alpha}L_{k}}[c_{l}]\right|^{2}.
\]
When $\alpha$ goes to zero, the term $e^{\frac{t}{2\alpha}L_{k}}[c_{l}]$
is the long time limit of the solution of (\ref{eq:perturbationannexeA}).
We use the nontrivial result for the case of non monotonous flows,
of \cite{Bouchet_Morita_2010PhyD}, that there exists a function $\omega_{d}^{\infty}(y)$
such that $\omega_{d}(y,t)\underset{t\rightarrow\infty}{\sim}\omega_{d}^{\infty}(y)e^{-ikUt}$
when there are no neutral modes. Hence $\left|e^{\frac{t}{2\alpha}L_{k}}[c_{l}]\right|\rightarrow|\omega_{d}^{\infty}(y)|$
, and the presence of the exponential in the integral ensures the
convergence of the whole. This proves that without neutral modes 
\[
2\alpha\left\langle |\omega|^{2}\right\rangle \underset{\alpha\rightarrow0}{\longrightarrow}|\omega_{d}^{\infty}|^{2}.
\]

How then can we compute the function $\omega_{d}^{\infty}$? We start
from equation (\ref{eq:perturbationannexeA}) that describes the linear
evolution of a perturbation $\omega(y,t)\mbox{e}^{ikx}$ of meridional
wave number $k$, and with streamfunction\textbf{ $\psi(y,t){\rm e}^{ikx}$.}
The idea is to transform equation (\ref{eq:perturbationannexeA})
into an inhomogeneous Rayleigh equation, as classically done, and
then to study its asymptotic solutions close to the real axis, which
is the limit $\delta\rightarrow0$, with the notations below. We introduce
the function $\varphi_{\delta}(y,c)$ which is the Laplace transform
of the stream function $\psi(y,t)$ i.e $\varphi_{\delta}(c):=\int_{0}^{\infty}{\rm d}t\psi(y,t)e^{-ik(c+i\delta)t}$
. The equation for $\varphi_{\delta}$ is
{\scriptsize 
\begin{equation}
\left(\frac{d^{2}}{dy^{2}}-k^{2}\right)\varphi_{\delta}(y,c)+\frac{\beta-U"(y)}{U(y)-c-i\delta}\varphi_{\delta}(y,c)=\frac{\omega(y,0)}{ik(U(y)-c-i\delta)},\label{eq:inhomogeneous_rayleigh}
\end{equation}}
with the boundary conditions that $\varphi_{\delta}$ vanishes at
infinity. We do not have a flow infinite in the $y$ direction, but
as was already stated, the properties of the flow become local for
large $K$. The choice to take vanishing boundary conditions at infinity
is done for convenience and it is expected that this particular choice
does not modify the physical behavior of the perturbation.

For all $\delta>0$ the function $\varphi_{\delta}$ is well defined.
The inhomogeneous Rayleigh equation (\ref{eq:inhomogeneous_rayleigh})
is singular for $\delta=0$ and for any critical point (or critical
layer) $y_{c}$ such that the zonal flow velocity is equal to the
phase speed: $U(y_{c})=c$\textbf{. }One can show that $\varphi_{\delta}$
has a limit denoted $\varphi_{+}$ when $\delta$ goes to zero. The
function $\omega_{d}^{\infty}$ is then given by
\begin{equation}
\omega_{d}^{\infty}(y)=ik(U"(y)-\beta)\varphi_{+}(y,U(y))+\omega(y,0),\label{eq:omegainfini}
\end{equation}
see \cite{Bouchet_Morita_2010PhyD} for more details.


\end{document}